\documentstyle[12pt]{article}

\begin{document}
\begin{titlepage}
\begin{center}
May, 1993      \hfill     MIT-CTP-2212 \\

\vskip 0.2 in
{\large Constraints on a Massive Dirac Neutrino Model
}
\vskip .2 in
       {\bf Thomas Wynter$^a$ }\vskip 0.3cm

         {\it and \\}
{\bf Lisa Randall$^{a,b}$}\footnotetext{$^a$ This work is supported
in part by funds provided by the U.S.
Department of Energy (DOE) under contract \#DE-AC02-76ER03069 and in
part by the Texas National Research Laboratory Commission
under grant \#RGFY92C6.\hfill\break}
\footnotetext{{$^b$} National
Science Foundation Young Investigator Award.\hfill\break
Alfred P.~Sloan
Foundation Research Fellowship.\hfill\break
Department of Energy Outstanding Junior
Investigator Award.}

        \vskip 0.3 cm
{\it Massachusetts Institute of Technology\\
Cambridge, MA 02139\\}

       \vskip 0.3 cm

\begin{abstract}
 We examine constraints on a simple neutrino
model in which there are three massless and three massive Dirac
neutrinos and in which the left handed neutrinos are linear
combinations of doublet and singlet neutrinos.  We
examine constraints from direct decays into heavy neutrinos,
indirect effects on electroweak parameters, and flavor
changing processes. We combine these constraints
to examine the allowed mass range for the  heavy neutrinos
of each of the three generations.
\end{abstract}
\end{center}

\end{titlepage}

\setlength{\baselineskip}{1.5\baselineskip}
\setcounter{page}{1}

\section {Introduction}
Many models of neutrinos have been proposed to accomodate
light or massless neutrinos.  In a model with no right
handed neutrinos, it is clear that neutrinos are massless.
However, if there exist additional states which can
play the role of Dirac partners to the left handed states,
it is perplexing why  neutrinos should be massless,
or at least much lighter than their charged counterparts.
Of course,
 neutrinos can be given small masses by
coupling them to the standard Higgs doublet
because of an extremely small Yukawa coupling,
but it is  more compelling to have
an explanation for their small mass.
A common explanation is the so called ``see--saw" mechanism,
in which the neutrinos remain light because
the additional right handed states have a large
Majorana mass. In such a model, neutrino
masses are naturally small, since they
are suppressed by the ratio of Dirac to Majorana masses,
which is generally taken to be small.

In this paper, we consider another viable alternative
(see for ex. [2],[3],[4]).  In addition to the three ``right handed"
neutrinos, there are three additional singlet particles.
A lepton symmetry is
imposed so that the only allowed mass terms are Dirac
masses coupling the right handed neutrino to
the standard left handed neutrino and to the
additional singlet states.  The conequence
is that there are three heavy Dirac neutrinos,
with mass determined primarily by the large
mass term connecting the singlet and right
handed neutrinos and three exactly massless neutrinos,
the states orthogonal to the massive ones.
Such a model has been considered before in several
contexts; most recently it has been considered
in the context of an Extended Technicolor Model
with a GIM mechanism [4]. In this type of model,
the additional neutrino states could be quite
light, on the order of a GeV.

However, there are many constraints on such neutrinos.
They are constrained
from direct searches for particles which have them in their
final state, by universality constraints, and by flavor changing constraints.
Cosmological arguments are often used to constrain neutrino
masses, but the neutrinos of this model are unstable so
they do not apply.
 In this
paper, we put these constraints together, making
reasonable assumptions on the form of the mass
matrix and mixing angles, to determine
the allowed parameter regime. Many of these constraints
apply quite generally to any model in which the left
handed neutrinos mix with singlet states.Similar  bounds were considered in
ref. [18].  This
paper updates the bounds, integrates them with those from LEP,
and incorporates flavor changing bounds.  We find
with reasonable assumptions described below, the
lightest neutrino can be as light as 2 GeV, although the
third generation neutrino should be much heavier, greater than 80 GeV.

The organization is as follows.
We first present the model
and we describe the approximations which we use to
reduce the parameter space.  We then consider constraints
from meson and $Z$ decays. Following this, we discuss
the constraints from the fact that $G_F$ will not
have the same relation to standard model parameters
when the muon cannot decay to the heavy neutrino state.
We then look at flavor changing processes, which
are in general permitted when no flavor symmetries are assumed.
However, we assume mixing angles similar to those of the standard
KM matrix, so there are approximate U(1) symmetries present.
We then put together the constraints and consider three
models which describe the ratio of masses of the heavy neutrinos
to determine the allowed parameter regime.
Finally, we conclude.

\section{The Model and Simplifying Assumptions}
Many models have incorporated the neutrino scenario
we discuss here. For example, it has been incorporated
into GUT models [2],[3]. More recently, it has been
shown how to incorporate such a model in an Extended
Technicolor scenario [4]. We only consider the pheomonology
of the lepton sector here, so we neglect the origin
of the model and focus on the neutrinos.

The standard model is  extended by introducing  three new left-handed
neutrinos $S_{L}$
 and three right-handed neutrinos $\nu_{R}$. Both left handed neutrinos
are coupled to the right handed neutrinos through Dirac matrices.
All other possible mass entries are forbidden by a lepton number symmetry.
Thus,
$$-{\cal L}_{mass}=\pmatrix{\nu_{R}&0\cr}\pmatrix{D&S\cr 0&0\cr}
                    \pmatrix{\nu_{L}\cr S_{L}\cr}+h.c.$$
This coupling  results in  three  massive
Dirac neutrinos and three massless
eigenstates. The mass matrices $D$ and $S$ have different mass scales.
The scale for $D$ is constrained by SU(2) symmetry breaking whereas
the scale for $S$ is not, so it is reasonable to expect the
masses in $S$ to be larger.

The mass of the heavy neutrinos is essentially determined by $S$.
The electron, muon and tau neutrinos are
a  superposition  of massless  and massive  eigenstates. The
mixing to the massive  neutrinos  will, however, be  small; it will be
of the  order  of
$M_{D}/M_{S}$,  where $M_{D}$ and $M_{S}$  are typical masses in  $D$
a
To see more  precisely how this mixing occurs,
we  need  to find the three massless eigenstates  $\nu^0$ as well as the
three  with mass  $\nu^H$.
The  mass matrix  can  be diagonalised by multiplying on  the left and the
right by unitary matrices:

\vspace{-.15 in}
\begin{center}
$\pmatrix{V&0\cr 0&0\cr}\pmatrix{D&S\cr 0&0\cr}${\huge U}
$=\pmatrix{0&0\cr 0&M}$\hspace{2ex} with \hspace{2ex}
$\pmatrix{\nu_L\cr S_L\cr}=${\huge U}$\pmatrix{\nu^0\cr\nu^H\cr}.$
\end{center}

The unitary matrix $V$ diagonalises $DD^{\dagger}+SS^{\dagger}$ to give $M^2$.
The  unitary matrix {\large U} is given by\\
\vspace{-.15 in}
\begin{center}
{\huge U}$=\pmatrix{U_{D}^{\dagger}\Lambda_{S}V'^{\dagger}&
  U_{D}^{\dagger}\Lambda_{D}V'^{\dagger}\cr
          U_{S}^{\dagger}\Lambda_{D}V'^{\dagger}&
 -U_{S}^{\dagger}\Lambda_{S}V'^{\dagger}\cr}$,
\end{center}
where the matrices  $V'$, $U_{D}$ and $U_{S}$  are unitary matrices.
They diagonalize $M^{-1}VD$ and $M^{-1}VS$ (where $M^{-1}$ is the
inverse  of the  diagonal mass matrix M) to give the diagonal matrices
$\Lambda_D$  and $\Lambda_S$:
$$V'^{\dagger}(M^{-1}VD)U_{D}^{\dagger}=\Lambda_D$$
$$V'^{\dagger}(M^{-1}VS)U_{S}^{\dagger}=\Lambda_S .$$

The fact that the same $V'$ appears on the
left for both these diagonalizations is a consequence of the fact that
the two matrix products
$M^{-1}VDD^{\dagger}V^{\dagger}M^{-1}$ and
$M^{-1}VSS^{\dagger}V^{\dagger}M^{-1}$ commute with each other,
which follows in turn from the fact that their sum is the unit matrix.
The most important part of {\large U} is
the top right 3 by 3 block $U_{D}^{\dagger}\Lambda_{S}V'^{\dagger}$
which links the electron,  muon and tau neutrinos to the massive
neutrinos  $\nu^H$.

To extract bounds on the mass scales of $S$ and $D$ we need to make
some simplifications to reduce the number of parameters.
We will make the simplification that the matrices $D$ and $S$ are
diagonalized by the same unitary matrices. In this case
$V'=I_{3\times3}$.
If we then redefine the fields $S_L$ by a unitary
transformation, absorbing the unitary matrix $U_S$, we can rewrite
the matrix  {\large U} as:\\
\vspace{-.15 in}
\begin{center}
{\huge U}$=\pmatrix{U_{D}^{\dagger}\Lambda_{S}& U_{D}^{\dagger}\Lambda_{D}\cr
          \Lambda_{D}& \Lambda_{S}\cr}$,
\end{center}

with $\Lambda_{D}^2+\Lambda_{S}^2=I_{3\times3}$.

In this model
the mass scale of the Dirac mass $S$ is assumed to be much higher
than the scale of the Dirac mass $D$. If this difference is sufficiently
large, we can make the further simplification that
$\Lambda_S=I_{3\times3}$. From here on the subscript {\em D} on
$\Lambda_{D}$ will be dropped.

 At this point, we still have a large number of parameters. We simplify by
assuming the matrix $U_D$ is similar in structure to
the KM matrix for quarks.
 We notice that if there were no singlet  left-handed neutrinos
$S_L$ the matrix $U_D$ would be the lepton equivalent of the KM matrix
in the quark sector. We take the individual elements to be of
the same magnitude as those in the KM matrix for quarks.

We will use these approximations from now on.
They leave six free parameters: $M_i$, the masses
of the heavy neutrinos and $M_{D_i}$, the masses induced by the mass
matrix $D$ which are defined as  $M_{D_i}=\Lambda_{i}\times M_i$.
In the following sections we will use experimental
results to put limits on these masses.

\section{Direct Searches for Heavy Neutrinos}
Many searches for massive neutrinos have already been
conducted. Massive neutrinos
have been sought in the decays of $\pi^+$ [5] [6] [7], $K^+$ [8] [9]
and charmed mesons [10] [11] [12], as well as in the neutral current
production of neutrino anti-neutrino pairs from $e^+\,e^-$
[13] [14] [15] collisions, and, more recently the decay of the $Z$ [16] [17].

\subsection{Meson Decays}
If it is kinematically allowed, any process involving the production
of neutrinos will be a source of heavy neutrinos. The creation
process, however, will be supressed since the weak eigenstate neutrinos
contain only a small mixing of the heavy neutrinos. Leptonic decays of
mesons are thus one place to look for heavy neutrinos.

At the lower end of the mass scale heavy neutrino creation in the
decay of $\pi^+$ mesons has been
investigated in references [5] [6] [7],
and those of $K^+$ mesons in references [8] [9].
These experiments
attempted to measure the mass of any heavy neutrino as it was created.
This was achieved by stopping the $\pi^+$ and $K^+$ mesons and
observing the energy of
positrons emitted in their decay. The method did not rely on any
assumptions about how the heavy neutrinos decayed. For massive
neutrinos with masses less than 300 Mev  these experiments placed
strict limits on the mixings, $|U_{ei}|^2$ and $|U_{\mu i}|^2$ of a
heavy neutrino $\nu^H_i$ into the electron and muon neutrinos.
For a range of masses
the matrix elements $|U_{ei}|^2$ and $|U_{\mu i}|^2$ were
constrained to be less than $10^{-5}$.
Since we assume the matrix $U_D$ is  almost diagonal,
we can get direct bounds on $\Lambda_1$ and $\Lambda_2$ of:
$\Lambda_1<3\times 10^{-3}$ for the mass range $35$ to $360$ $MeV$,
and $\Lambda_2<3\times 10^{-3}$ for the mass range $80$ to $325$
$MeV$ (see Figs(1) and (2)).

Further limits on $|U_{ei}|^2$ and $|U_{\mu i}|^2$ come from the decays of
charmed $D$ mesons, [10] [11] [12], (see Figs(1) and (2)).
Although similar searches can in principle be performed with
decays of the $B$, they have not yet been done. In total the
limits from meson decays, with a few gaps, restrict
$\Lambda_1<3\times 10^{-3}$ for the mass range $35$ $MeV$ to $2$ $GeV$,
and $\Lambda_2<3\times 10^{-3}$ for the mass range $80$ $MeV$ to $2$
$GeV$. For much of these ranges the bounds are much stricter than this.

\subsection{$e^+\,e^-$ Collisions at Low CM Energy}
Heavy neutrino anti-neutrino pairs would be created by weak
interaction currents in $e^+\,e^-$ annihilations, but since the the center of
mass
energy of these collisions is less than the $W$ and $Z$ mass the cross
section for the
creation of these neutrinos is extremely small.
Experiments [13] [14] [15] were aimed at detecting the decay of
a heavy fourth generation neutrino and they thus made the assumption that
the heavy neutrino had the same coupling to the $Z$ and $W$ as the
other neutrinos. This is not the case for the model studied in this
paper where each heavy neutrino introduces a mixing angle factor of
$|U_{li}|^2$ into the weak interaction couplings. Reinterpreting the
data of these experiments including the extra mixing angles results in
constraints that are negligible in comparison to the other bounds studied in
this paper.

\subsection{$Z$ decays}
Massive neutrinos, lighter than $M_Z$, would also be created in $Z$
decays, and the experiments [16], [17] have already
conducted searches for heavy isosinglet neutrinos; the type discussed
in this paper.
The most abundant supply of heavy neutrinos would come from the
decay of a $Z$ into one heavy neutrino and one massless neutrino.
For a $Z$ decaying into
a heavy neutrino $\nu^{H}_i$ (lighter than the $Z$) and any of the
three massless anti-neutrinos
$\nu^{0}_{j}$ the creation is suppressed by:
$$R_i=\sum_{j=1}^3\Bigl|\sum_{l=e,\mu,\tau}{\rm U}_{jl}^{\dagger}
      {\rm U}_{l(i+3)}\Bigr|^2
          =\Lambda_{i}^2 ,$$
where $R_{\nu_i}$ has the
following meaning: if $N$ is the number of neutrinos (from one
family of the standard model) created in the experiment then then
number of heavy neutrinos, $\nu_i^H$, created is $R_iN$.

Experiments aim to detect the neutrino by its decay.
The decay of the neutrino would be quite distinctive. In general, it
will decay to a high energy lepton and a virtual $W$ or $Z$, which
would then decay into leptons, or hadrons if the neutrino is massive
enough.  The total decay rate can be written in terms of the rate for muon
decay as follows[18]:

$$\Gamma(\nu_i^H\rightarrow\,leptons/hadrons)=
            \sum_l
           \bigl|{\rm U}_{l(i+3)}\bigr|^2
             \Bigl({M_i\over M_{\mu}}\Bigr)^5\Phi_l(M_i)
           \Gamma(\mu\rightarrow e\nu\bar{\nu})$$
where $\Phi_l(M_i)$ is a factor that weights the decay rate for a
single channel by the effective number of
channels into which there is sufficient energy to decay and takes
into account the different  Feynman diagrams.

There are two reasons why decays like this might not have been seen in
experiments:
\begin{description}
\item [Very few heavy neutrinos are produced.] If we assume that
nearly all the neutrinos decay inside the detector, then the fraction, $R_d$,
that decay inside the detector is given by:
$$R_d=R_i=\Lambda_i^2.$$
If $R_i$ is sufficiently small, no neutrinos would be detected.

\item [The neutrinos have a long lifetime.] If the neutrinos are
light, they could have a very long lifetime and thus decay
almost entirely outside of the detector.
\end{description}

We can then calculate the
fraction, $R_d$ of $Z$'s that would decay inside the detector:
$$R_d=R_i\Bigl(1-\exp\bigl({S_d\over c}\gamma^{-1}
          \Gamma(\nu_i^H\rightarrow\,leptons/hadrons)\bigr)\Bigr),$$
where $\gamma$ is the time dilation factor due to the
relativistic motion of the neutrino, and for $M_Z>>M_i$ is given by
$\gamma=M_Z/ 2M_i$, $S_d$ is the size of the detector
and $c$ is the speed of light.

The experiment of reference [16] involved a search through $4\times10^5$
hadronic $Z$ decays and placed limits of $\Lambda_i<0.014$ over the
range $5$ to $50$ $GeV$. Above $50$ $GeV$ the phase space for heavy
neutrino production becomes smaller and the limits placed on the
$\Lambda_i$ become less strict. Below $5$ $GeV$ the limits are reduced
due to the long lifetime of the neutrinos.
In Figs (1), (2) and (3), are marked  out the forbidden regions in the $M_i$,
${1\over\Lambda_i}$ plane for the three heavy neutrinos.

\section{Changes In Weak Interaction Decays and Parameters}

Aside from direct searches for the heavy neutrino, the
existence of the heavy neutrino will affect precision
measurements of various electroweak processes.
This can be the case because $G_F$ will no longer
have the standard model relation to $\sin \theta_W$,
since the muon decay rate will be different if the
muon cannot decay into the heavy neutrino. This would
change the relation between precisely measured electroweak
parameters, for example the $W$ mass or $\sin^2 \theta_W$ as
measured in the forward--backward asymmetry.  Furthermore,
it would lead to an apparently nonunitary KM matrix.

Further constraints come from pion decay branching ratios
if the heavy electron or muon neutrinos are heavier than
the pion. Similarly, universality could be violated and
would be seen in tau decay.  Finally, the $Z$ width
can be affected, both indirectly though a change in
the extracted $\sin^2 \theta$, and directly if the
neutrinos are heavier than the $Z$.

\subsection{Muon decays and the Fermi coupling constant $G_F$}

The Fermi coupling constant $G_F$ is the effective coupling constant for four
fermi interactions and is measured extremely accurately
from muon decays.
If the mass of the massive neutrinos is
greater than that of the muon, the decay width for the muon would be
decreased,
since it would not be able to decay into the heavy neutrinos; this in
turn would lead to a change in the predicted value of $G_F$.

Specifically:
$$(G_F)^2_{new}=\sum_{i,j=1}^3|U_{e\,i}U_{j\,\mu}^{\dagger}|^2
                                          (G_F)^2_{old},$$
which leads to
$$\delta(G_F)^2_{muon\,decays}
                   \simeq-(\Lambda_{1}^2+\Lambda_{2}^2)$$
where $\delta$ means the fractional change. Of course $G_F$ is
a measured number. What we mean here is the change in the coefficient
of the four fermion operator which yields muon decay.

\subsection{Semileptonic Decays and the KM Matrix}
The estimates of the semileptonic processes would also be affected but to
a lesser extent. The same value of $G_F$ is also used for the
effective coupling constant for
semileptonic decays, where elements of the KM matrix
are determined. One would expect  these elements
to be part of a unitary matrix.

The important point to consider is that the effective coupling
constants for the leptonic and semileptonic four fermion interactions
would no longer be the same, and if it was assumed that they were, the
predicted matrix elements for the KM matrix would no longer be those
of a unitary matrix. We can check the unitarity of the KM matrix by
looking at the matrix elements $(KM)_{ud}$, $(KM)_{us}$ and $(KM)_{ub}$; the
sum of their square magnitudes must add up to one. The effect of
having heavy neutrinos would be to make this sum slightly bigger than
one. The most important
shift will come from the change in nuclear beta decays used to
determine the $(KM)_{ud}$ element. Consequently, what must be compared
are the
changes in the value of $G_F$ and  in the rates for nuclear beta decays.

Specifically, as above,
$$(G_F)^2_{new}=\sum_{i,j=1}^3|U_{e\,i}U_{j\,\mu}^{\dagger}|^2
                                          (G_F)^2_{old},$$
which leads to
$$\delta(G_F)^2_{muon\,decays}
                   \simeq-(\Lambda_{1}^2+\Lambda_{2}^2)$$
and similarly
$$\delta(beta\,decay)\simeq-\Lambda_{1}^2$$
where $\delta$ means the fractional change.
The fractional change in the width of the muon minus the fractional change
in nuclear beta decays must be less
than the experimental uncertainty in the sum of the matrix elements.
 From reference [19]:
$$|(KM)_{ud}|^2+|(KM)_{us}|^2+|(KM)_{ub}|^2=1\,(+8.6\times 10^{-4},\,
-4.7\times 10^{-3}).$$
This leads to $\Lambda_{2}<6\times 10^{-2}$ $(2\sigma)$ if the
neutrinos are heavier
than the muon.

\subsection{$M_W$ and $\sin\theta_W$}
Changes in $G_F$ would also affect the prediction of other weak
interaction parameters. The ratio of the mass of the $W$ and $Z$ for
example depends upon $G_F$. Specifically [19]:
$${M_W^2\over M_Z^2}={1\over2}\Bigl(1+\bigl(1-{4\pi\alpha(1+\delta v)\over
                                   \sqrt2M_Z^2G_F}\bigr)^{1\over2}\Bigr),$$
where $\delta v$ is a radiative correction parameter much less than
one. Using the above the change in the predicted value of $M_W/M_Z$
due to the change in $G_F$ is:
$$\delta({M_W/M_Z})=0.088\times\delta(G_F).$$
Current experimental bounds [20] place
$\delta(M_W/M_Z)<7.7\times 10^{-3}$ $(2\sigma)$
which gives a bound for $G_F$ of:
$$\delta(G_F)<8.8\times 10^{-2}\,\,\,(2\sigma)$$
In fact this bound is too strong due to the uncertainty in the
top quark mass. However, since it iss less strict than the bound
coming from the KM matrix, it will not be incorporated.

The Weinberg angle  $\sin^2\theta_W$ also depends on $G_F$ (the on
shell definition is: $\sin^2\theta_W=(1-M_W^2/M_Z^2)$ and this can be
compared with the forward backward assymetry of the process
$e^+\,e^-\rightarrow f\, \bar{f}$, which depends on $\theta_W$.
However, this too is weaker than the constraint from the unitarity
of the KM matrix.

\subsection{Pion Decay Branching Ratios}
The ratio of the two decay channels for a $\pi^{\pm}$;
$\pi\rightarrow e\,\nu_e$ and $\pi\rightarrow \mu\,\nu_{\mu}$, provides
another bound [21]. In this model:
$${\Gamma(\pi\rightarrow e\,\nu_e)\over\Gamma(\pi\rightarrow\mu\,\nu_{\mu})}
    =1.233\times 10^{-4}\bigl(1-\sum_{i=1}^3(|U_{ei}|^2-|U_{\mu i}|^2)\bigr),$$
where the factor $1.233\times 10^{-4}$ is the theoretical value of the
ratio in the standard model[22]. The mixing angle factors apply for
neutrinos too heavy to be
produced. Experimentally the ratio is known to be:
$(1.218\pm0.014)\times 10^{-4}$ [23]. Using the fact that $U_D$ is
almost diagonal and that $\Lambda_2<6\times 10^{-2}$ leads to:
$$0.035>(\Lambda_1^2-\Lambda_2^2)\,\,\Longrightarrow\Lambda_1<0.18
       \,\,\,(2\sigma)$$

\subsection{Tau decays}
If the neutrinos are all heavier than the tau then the decay width of
the tau would also be affected. As for the case of
the muon decay it can be shown that the partial width
$\Gamma(\tau\rightarrow e\nu_{\tau}\nu_e)$ would be reduced by
$\sim(\Lambda_{1}^2+\Lambda_{3}^2)$  and the partial width
$\Gamma(\tau\rightarrow \mu\nu_{\tau}\nu_{mu})$ would be reduced by
$\sim(\Lambda_{2}^2+\Lambda_{3}^2)$. Consequently, the partial
width for
decay into leptons would be reduced by:
$$\sim({1\over 2}\Lambda_{1}^2+{1\over 2}\Lambda_{2}^2+\Lambda_{3}^2).$$
This fractional change minus the fractional change in the width for
muon decay must be less than the experimental uncertainty of the
partial width for the tau. This gives a further bound on the
$\Lambda_{i}$:
$$\bigr|{1\over 2}\Lambda_{1}^2+
    {1\over 2}\Lambda_{2}^2-\Lambda_{3}^2\bigl|\leq 0.015\,\,\,(1\sigma)$$
where the uncertainty in the partial width of the tau is 1.5\% [19].
To obtain a bound for $\Lambda_3^2$ we use the
following formula for calculating the error at the $1\sigma$ level of a
sum of terms each with their own errors:
$$\Lambda_3^2=\bigl( \bigl({1\over2}\Lambda_1^2\bigr)^2_{1\sigma} +
\bigl({1\over2}\Lambda_2^2\bigr)^2_{1\sigma} + 0.015^2\Bigr)^{1\over2},$$
where, at the $1\sigma$ level,
we use the bounds from the previous section
 $\Lambda_1^2<0.017$ and $\Lambda_2^2<0.0018$.
This leads to a bound on $\Lambda_3$ at the $2\sigma$ level of:
$$\Lambda_{3}<0.18\,\,\,(2\sigma),$$

\subsection{The Width of the $Z$}
For neutrinos heavier than the $Z$ the width will be reduced, since the
decay into the heavy neutrinos will no  longer be kinematically
allowed. Experimentally, the partial width,$\Gamma_{\nu\bar{\nu}}$ of
the $Z$ is known to an accuracy of $1.8\%$ [24]. In this model
$$\Gamma_{\nu\bar{\nu}}\propto {1\over 3}\sum_i|1-\Lambda_i^2|
                                           =1-{1\over 3}\sum_i\Lambda_i^2,$$
where the sum over $i$ is only over neutrinos heavier than the $Z$.
This gives the bound:
$$\sum_i\Lambda_i^2<0.108\,\,\,\,(2\sigma).$$
There is also an effect if muon decay is changed. However it is not
numerically as important.

In Figs (1), (2) and (3) are plots of the bounds placed on the
$\Lambda_i$ by all the processes considered in sections four and five.

\section{Lepton Flavor Changing Processes}
Flavor changing processes were also examined in this model. These
processes are exactly analagous to flavor changing
processes in the quark sector. Three processes with strong
experimental bounds were considered [19]:

(i) $\mu\rightarrow e \gamma$; experimentally:
        ${\Gamma (\mu\rightarrow e \gamma)\over
           \Gamma (\mu\rightarrow e \nu\bar{\nu})}< 5\times 10^{-11}$,
\\

(ii) $\mu\rightarrow e e^+ e^-$; experimentally:
        ${\Gamma (\mu\rightarrow e e^+ e^-)\over
           \Gamma (\mu\rightarrow e \nu\bar{\nu})}< 10^{-12}$, and
\\

(iii) $\mu\, Ti\rightarrow e\,Ti$; experimentally:
        ${\Gamma (\mu\, Ti\rightarrow e\,Ti)\over
           \Gamma (\mu^-\,Ti\,capture)}< 5\times 10^{-12}$.\\
\\
These processes can only occur via loop diagrams involving the
exchange of virtual neutrinos. The  couplings of the neutrinos  to the
muon and the electron involve the unitary matrix {\large U}; specifically,
the neutrino-$W$-muon vertex includes a factor {\large
U}$_{i\mu}^{\dagger}$ for coupling to the $i$th neutrino, and the factor
{\large U}$_{ei}$ is included with the neutrino-$W$-electron
vertex. The amplitudes are obtained by summing over all intermediate states
$i$. All terms proportional to the sum $\sum_{i=1}^6${\large
U}$_{ei}${\large U}$_{i\mu}^{\dagger}$ are automatically  cancelled
since {\large  U} is unitary.  This is  the GIM mechanism. It is an
analog of the strong suppression of neutral current flavor
changing processes in the quark sector.  Notice this is independent
of the approximations we made.

In all cases there is very strong  GIM suppression.
Flavor changing processes can only proceed via an intermediate
heavy neutrino state. But the coupling to
the charged neutrinos is then suppressed. Therefore these
constraints will only dominate in the region where there are no other strong
constraints, namely for neutrinos more massive than the $Z$.

For the purpose of calculations, the masses of the electron,
muon and tau and the mass matrix $D$ are generated in the standard way by
coupling to the Higgs, so that the loops  involved charged Higgs.
t'Hooft gauge is used throughout, simplifying the form of the propagators and
setting the masses of the $W$ and the charged Higgs to be the same.
We will now consider, in detail, the three flavor changing processes.

\subsection{ $\mu\rightarrow e \gamma$}

A calculation for the case where the neutrino masses are much less
than $M_W$ is elaborated in Cheng and Lee [25].
Below we will give an outline of the
calculation  for the general case, where the neutrino masses are not assumed
to be less than $M_W$. To simplify the
calculations the electron is taken to be massless. The general form of
the amplitude is constrained by gauge invariance, thus the
gauge invariant form of the amplitude with a massless electron is
given by
$$<e,\gamma |(S-1)| \mu>=
  A\bar{u}_e\bigl[(1-\gamma_5)ik^{\nu}\epsilon^{\lambda}
       \sigma_{\lambda\nu}i\slash\hspace{-1.2ex}{\partial}\bigr]u_{\mu},$$
where $k$ is the photon four momentum, $\epsilon$ is the polarization  of
the photon and $A$ is a constant to be
determined.  The partial derivative term is
included to ensure that, in the hypothetical case where the muon mass goes
to zero, only the lefthanded component of the muon coupled to the
$W$ is involved in the decay. The amplitude can then be rewritten using
the Gordon decompostion as:
$$<e,\gamma |(S-1)| \mu>=A\,m_{\mu}\bar{u}_e\bigl[(1+\gamma_5)(2\epsilon .p
                       -m_{\mu}\slash\hspace{-1.2ex}{\epsilon})\bigr]u_{\mu},$$
where $m_{\mu}$ is the muon mass and $p$ is the four momentum of
the incoming muon.
To simplify the calculation, only the terms proportional to $\epsilon .
p$ need to be calculated. In principle there are 8 possible diagrams
contributing (see Fig.(8a)). Diagrams $4$ to $8$ contain only the
$\slash\hspace{-1.2ex}{\epsilon}$ term and thus can be ignored. They will
cancel with similar terms coming from the first four diagrams.

In  evaluating diagrams 1 to 4 we can define a factor $I_{j}^{i}$
for each of the diagrams $i=$1 to 4, and for each of the neutrinos
$j=$1 to 6. Summing over the six neutrinos $j$ the contribution of diagram $i$
to the
constant $A$ is
$$-iK\sum_{j=1}^6 U_{ej}U_{j\mu}^{\dagger}I_{j}^{i}/(32 \pi^2 M_{W}^2)$$
where $K=-e^3/(2\sin^2\theta_W)$. We can also  define the sum
$$I_{\mu e}=\sum_{i=1}^{4}\sum_{j=1}^6 U_{ej}U_{j\mu}^{\dagger}I_{j}^{i}$$
so that the total contribution to the constant $A$ from all the diagrams is
$$-iKI_{\mu e}/(32 \pi^2 M_{W}^2).$$
Performing the calculations gives
the following results for the $I_{j}^{i}$ :
\begin{eqnarray*}
I_{j}^{1}=&
   -\bigl [a_j(a_{j}^2-3a_j+{31\over12})+
     a_{j}^3(a_{j}-{3\over2})\delta_{j}^2\log\delta_j\bigr ],\\
I_{j}^{2}=&
   -\bigl [{1\over2}a_{j}^3-{1\over4}a_{j}^2-{7\over12}a_j+{1\over3}+
     {1\over2}a_{j}^3(a_j+1)\delta_{j}^{2} log\delta_j\bigr ],\\
I_{j}^{3}=&0,\,\mbox{and}\\
I_{j}^{4}=&
   \bigl [{1\over2}a_{j}^2-{3\over4}a_{j}+
      {1\over2}a_{j}^3 \delta_{j}^2\log\delta_j\bigr ],
\end{eqnarray*}
where the variable $\delta_j$ is given by $\delta_j=0$ for $j=1,2,3$
and by $\delta_j=(M_{\nu_(j-3)}/M_W)^2$  for $j=4,5,6$, and  $a_j$
is defined as
$a_j=1/(1-\delta_j)$. Using the expression for {\large U},  the sum
$I_{\mu e}$ is then given by:
$$I_{\mu e}=\sum_{j=4}^6 U_{D_{e(j-3)}}U_{D_{(j-3)\mu}}^{\dagger}
\Lambda_{(j-3)}^2$$
 $$\times \bigl [-{3\over2}a_{j}^3+{15\over4}a_{j}^2-{11\over4}a_{j}+{1\over2}
       -{3\over2}a_{j}^4 \delta_{j}^3\log\delta_j\bigr ].$$
This can be approximated for the two cases where the neutrino masses
are all either much less than or much
greater than the mass of the $W$. Thus,
\begin{eqnarray*}
I_{\mu e}\simeq& -\sum_{j=1}^3 U_{D_{ej}}U_{D_{j\mu}}^{\dagger}
             \Lambda_{j}^2{1\over4}\delta_{(j+3)}
                   \,\,\,\mbox{if all}\,M_j<M_W\\
         \simeq& \sum_{j=1}^3 U_{D_{ej}}U_{D_{j\mu}}^{\dagger}
               \Lambda_{j}^2
  ({1\over2}+{1\over\delta_{(j+3)}}{11\over4}+{3\over2}\log\delta_{(j+3)})
                   \,\,\,\mbox{if all}\,M_j>M_W.
\end{eqnarray*}
Averaging over the initial spins and summing over the final spins and
momenta leads to the decay rate:
$$\Gamma_{\mu\rightarrow e\gamma}={\alpha G_{F}^2 m_{\mu}^5 \over
                                            128\pi^4}|I_{\mu e}|^2,$$
where $\alpha$ is the fine structure constant.
This can be compared to the decay rate
$\Gamma_{\mu\rightarrow e\gamma}={G_{F}^2 m_{\mu}^5 \over
          192\pi^3}$ to get the ratio which can then be compared to
experiment to get an upper bound on $I_{\mu e}$:
$${\Gamma (\mu\rightarrow e \gamma)\over\Gamma (\mu\rightarrow e
   \nu\bar{\nu})} ={3\alpha\over 2\pi}|I_{\mu e}|^2< 10^{-10}\,\,\,(2\sigma),$$
$$\Longrightarrow |I_{\mu e}|<1.7\times 10^{-4	}\,\,\,(2\sigma).$$

\subsection{ $\mu\rightarrow e e^+ e^-$}

This process can occur via extensions of the diagrams in
$\mu\rightarrow e \gamma$ where the $\gamma$ is virtual and
splits into an electron positron pair, or it can take place via box
diagrams (see Fig (8b)). The contribution from box
diagram(1) dominates the other box diagrams. This is
due to the fact that the other diagrams
involve  exchange of virtual charged Higgs whose coupling to the
electron and muon is supressed by a factor of the order of the Dirac
mass $D$ over $M_W$. The contribution from the extensions of the
diagrams in $\mu\rightarrow e \gamma$ is of the same order as
the first box diagram, but for an order of magnitude estimate we
approximated the whole amplitude from the first box diagram only,
(diagram(1) Fig(8b)).

The amplitude for the box diagram is calculated using the
approximations that the neutrino masses are much greater than the
muon mass, and that the electron mass is zero. Using these approximations
the amplitude for the process is:
$${\cal A}=-{ie^4 \over 32\pi^2\sin^4\theta_W M_{W}^2}
              \bar{u}_{q_1}P^-\gamma_{\eta}u_p
                  \bar{u}_{q_2}P^-\gamma^{\eta}v_{q_3}J_{\mu e},$$
where the dimensionless factor $J_{\mu e}$ is:
$$J_{\mu e}= \sum_{j=1}^6 U_{ej}U_{j\mu}^{\dagger}
 \Lambda_{(j-3)}^2 (a_j-1+a_{j}^2 \delta_j \log\delta_j )$$
with the variable $\delta_j$ given by $\delta_j=0$  for $j=1,2,3$ and by
$\delta_j=(M_{\nu_(j-3)}/M_W)^2$ for $j=4,5,6$,
and  $a_j$ defined as $a_j=1/(1-\delta_j)$. As for the previous
flavor changing process, we can make approximations for the cases
where all the neutrino masses are less than the mass of the $W$ and
where they are all much greater than the mass of the $W$. We obtain:
\begin{eqnarray*}
J_{\mu e}\simeq& \sum_{j=1}^3 U_{D_{ej}}U_{D_{j\mu}}^{\dagger}
             \Lambda_{j}^2\delta_{(j+3)}(1+\log\delta_{(j+3)})
               \,\,\,\mbox{if all}\,M_i<M_W\\
          \simeq& \sum_{j=1}^3 U_{D_{ej}}U_{D_{j\mu}}^{\dagger}
               \Lambda_{j}^2
             -[1+{1\over\delta_{(j+3)}}(1-\log\delta_{(j+3)})]
                \,\,\,\mbox{if all}\,M_i>M_W.\\
\end{eqnarray*}
Averaging $|{\cal A}|^2$ over the initial spin states and summing over
the final spins and momentums leads to the decay rate:
$$\Gamma_{\mu \rightarrow e e^+ e^-}={G_{F}^2 m_{\mu}^5 \alpha^2
                        \over 768\pi^5\sin^4\theta_W}|J_{\mu e}|^2,$$
where $\alpha$ is the fine structure constant.
As before this can be compared to the decay rate
$\Gamma_{\mu\rightarrow e \nu\bar{\nu}}={G_{F}^2 m_{\mu}^5 \over
          192\pi^3}$ to obtain the ratio which can then be compared to
experiment to get an upper bound on $J_{\mu e}$:
$${\Gamma (\mu\rightarrow e e^+ e^-)\over\Gamma (\mu\rightarrow e
   \nu\bar{\nu})} ={\alpha^2\over 4\pi^2\sin^4\theta_W}|J_{\mu e}|^2
                             <  2\times 10^{-12}\,\,\,(2\sigma),$$
$$\Longrightarrow |J_{\mu e}|<2.8\times 10^{-4}\,\,\,(2\sigma).$$

\subsection{ $\mu\, Ti\rightarrow e\,Ti$}

O.Shanker [26] has performed some careful calculations for $\mu\,e$
conversion for different nuclei. These calculations involve using an
effective Hamiltonian for the muon-electron-q-q vertex where the q's
represent either two up quarks or two down quarks. This effective
Hamiltonian is obtained from box diagrams very similar to the ones in
the previous section except that the outgoing electrons, labeled with
momenta $q_2$ and $q_3$, are replaced by an incoming and outgoing
quark in the Titanium nucleus, see Fig(8c). The calculation for the
amplitude from the previous section can be carried over with very
few changes to give the amplitude for the process involving the up
quark; $\mu u\,\rightarrow\,e\,u$:
$$A_{\mu u\,\rightarrow\,e\,u}=-i{G_F\alpha J_{\mu e}\over
    \sqrt{2} 4\pi \sin^2\theta_W}\bar{e}_{q_1}P^-\gamma^{\lambda}
      \mu_p \bar{u}_{q-3}P^-\gamma_{\lambda}u_{q_2},$$
and the same amplitude for the down quark. These calculations assume that
the quark mass is less than the neutrino mass.
We can then write an effective Hamiltonian for this interaction:
$$H_{eff}=g{2G_F\over sqrt{2}}\bar{e}P^-\gamma^{\lambda}\mu
           (\bar{u}P^-\gamma_{\lambda}u+\bar{d}P^-\gamma_{\lambda}d),$$
where $g= \alpha J_{\mu e}/(4\pi \sin^2\theta_W)$.
Using the calculations of O.Shanker [26], we can then obtain the ratio
between the decay rate for $\mu\, Ti\rightarrow e\,Ti$ to the rate for
muon capture by the nucleus which can be compared to experiment to
obtain another bound on $J_{\mu e}$:
$${\Gamma (\mu\, Ti\rightarrow e\,Ti)\over\Gamma (\mu^-\,Ti\,capture)}
            =265.64{\alpha^2 \over 16\pi^2 \sin^4\theta_W}|J_{\mu e}|^2\leq
              10^{-11}\,\,\,(2\sigma)$$
$$\Longrightarrow |J_{\mu e}|<7.6 \times 10^{-5}\,\,\,(2\sigma).$$

The diagrams for the two processes, $\mu\, Ti\rightarrow e\,Ti$ and
muon capture, are essentially the same
as the diagrams for $\mu\rightarrow e e^+ e^-$ and
$\mu\rightarrow e\nu\bar{\nu}$ but the ratio of the decay rates
of the first two processes is much greater than for the second two.
Thus although the bounds placed by experiment on
$\mu\, Ti\rightarrow e\,Ti$ are not as strong as those for
$\mu\rightarrow e e^+ e^-$, it is the process $\mu\, Ti\rightarrow e\,Ti$
which places the strongest bounds on the size of $J_{\mu e}$.

This difference can be explained by coherence
effects. The dominant process for $\mu\, Ti\rightarrow e\,Ti$
leaves the $Ti$ nucleus in it's ground state [26], which is a coherent
process involving summing the amplitude  over all the nucleons.
Muon capture, on the other hand, is an
incoherent process involving summing the square of the amplitude over all
the protons.

\section{Discussion and Conclusions}
One of the desired results of this model was that it would provide a
scenario in
which weak interaction symmetry breaking could give the neutrinos a
mass matrix on a scale similar to that of the electron, muon and tau,
while still maintaining massless neutrinos. To investigate this all
the plots discussed in this section are marked with a dashed line
along which the masses $M_{D_i}$ induced
by the mass matrix $D$ (from weak interaction symmetry breaking) are
the same as the electron, muon and tau. For all the plots the excluded
regions lie to the left of the curves.

The bounds from sections three and four are plotted separately for each
of the $\Lambda_i$ in Figs (1), (2) and (3). To fulfill the scenario
in which $M_{D_1}=M_e$, $M_{D_2}=M_{\mu}$ and $M_{D_3}=M_{\tau}$ we
see that the mass of the third neutrino must be greater than $M_W$,
the second must be heavier than $10$ $GeV$ and the first heavier than
$2$ $GeV$.

To examine if there are further restrictions from the flavor changing
processes of section five, the $\Lambda_i$ have to be plotted on the
same graph since the factors $I_{\mu e}$ and $J_{\mu e}$ are functions
of all three $\Lambda_i$. In fact due to the very small mixing of the
third massive neutrino $\nu^H_i$ into the electron and tau neutrinos
(the mixing matrix $U_D$ is chosen in this analysis to be like the KM matrix)
$I_{\mu e}$ and $J_{\mu e}$ are virtually independant of $\Lambda_3$.

Figs (4), (5), and  (6)  plot out the constraints from the flavor
changing processes for three different scenarios. They all assumed that the
masses $M_{D_i}$ generated by the mass matrix $D$ were in the same
ratio as the masses of the electron muon and  tau ie.
$M_{D_1}:M_{D_2}:M_3=M_e:M_{\mu}:M_{\tau}$. The dotted line, as
before, marks out the line along which the masses $M_{D_i}$ are
actually the same as the electron, muon and tau masses. The flavor
changing processes are plotted alongside all the other constraints
from sections three and four. It is immediately clear that flavor
changing processes do not rule out any of the line along which $M_{D_1}=M_e$,
$M_{D_2}=M_{\mu}$ and $M_{D_3}=M_{\tau}$. The bounds from $Z$ decays
and, for the lightest neutrino, meson decays are much more important.

\subsection{Three scenarios}
In the scenarios that follow four different ratios of the neutrino
masses $M_i$ are considered. The bounds given at the end of the
discussion of each scenario assume that $M_{D_1}=M_e$,
$M_{D_2}=M_{\mu}$ and $M_3=M_{\tau}$.

\begin{description}
\item[Scenario 1] Fig(4) $M_1=M_2=M_3$. \newline
In Fig(4) are plots of the allowed regions taking into account all
the experimental constraints from sections three, four and five. Areas to the
left of the curves are ruled out. The plot is of the
mass $M_2$ of the second heavy neutrino against ${1\over\Lambda}$, the
ratio between the two mass scales generated by $S$ and $D$. The most
important constraint comes from the limits set by $Z$ decays on the
third neutrino. If they are to lie on the dashed line all three
neutrino masses are constrained to be greater than $M_W$.

\item[Scenario 2] Fig(5) $M_1:M_2:M_3=1:15:60$. \newline
Again, areas to the left are ruled out by
experiment and the plot is of the
mass $M_2$ of the heaviest neutrino against ${1\over\Lambda_{2}}$, the
ratio between the masses $M_2$ and $M_{D_2}$. In this case the most
important constraints are those set by D decays on the mass of the
first neutrino and those set by $Z$ decays on the mass of the third neutrino.
If the masses are to lie on the dashed line $M_2$ must be greater than $30$
$GeV$. Dividing this by $15$ and multiplying by $4$ gives the bounds for
the first and third neutrinos respectively. The bounds for the three
neutrinos are thus: $M_1>2\,GeV$ which is equivalent to $M_2>\,30GeV$
and $M_3>120\,GeV$.

\item[Scenario 3] Fig(6) $M_1:M_2:M_3=M_e:M_{\mu}:M_{\tau}$\newline
In this scenario it is the constraints set by D decays on the mass of
the first neutrino that are most important and the corresponding
bounds for the three masses are (for masses lying on the dashed line):
$M_1>2\,GeV$, $M_2>\,400\,GeV$ and $M_3>\,3500GeV$.

\end{description}

\subsection{Conclusions}
In this paper we have examined the experimental consequences of a
model of massive neutrinos and have excuded a large
region of the parameter space. Specifically we have found that, in the
scenario where the mass contributions, $M_{D_i}$, from weak interaction
symmetry breaking are the same as those for the electron, muon and tau,
the neutrino masses are  approximately constrained as follows:
$$M_1>2\,GeV,\,\,\,\,\,M_2>10\,GeV,\,\,{\rm and}\,\,M_3>80\,GeV.$$
This means that either the Dirac mass connecting standard left handed
neutrinos to right handed neutrinos has entries less than
their charged counterparts, or one would not expect all the neutrinos
to be light.  It is clearly nonetheless of interest to improve
the bounds.  Clearly improved statistics at LEP will give stronger
constraints.  Furthermore, the bound on $M_1$ can be improved
by looking for heavy neutrinos in $B$ decays.

\section*{Acknowlegements}
 L. R. thanks  CERN and the Rutgers University physics department
for their hospitality while this work was being completed.

\section*{References}
$[1]$ H. Harari \& Y. Nir, Nucl . Phys . B 292 251-297 (1987).\\
$[2]$ C.N. Leung, J.L.Rosner, Phys . Rev. D 28 2205 (1983).\\
$[3]$ D. Wyler, L. Wolfenstein, Nucl . Phys . B 218 205 (1983).\\
$[4]$ L. Randall, MIT-CTP-2112, Bull. Bd.: hep-ph@xxx.lanl.gov-9211268
(1992). \\
$[5]$ N. De Leener-Rosier et al., Phys . Rev. D 43  3611 (1991).\\
$[6]$ G. Azuelos et al., Phys . Rev. Lett. 56 2241 (1986).\\
$[7]$ D.A. Bryman et al., Phys . Rev. Lett. 50 1546 (1983).\\
$[8]$ T. Yamazaki et al., Proceedings of the Eleventh International
conference on Neutrinos and Astrophysics at Dortmond (World
scientific, Singapore, 1984) p183 (1983).\\
$[9]$ J. Heintz et al., Nucl . Phys . B 149 365 (1979).\\
$[10]$ J. Dorennbosch et al., Phys . Rev. Lett. B 166 473 (1986).\\
$[11]$ F.J. Gilman et al., Phys . Rev. D 32 324 (1985).\\
$[12]$ F. Bergsma et al., Phys . Lett. B 128 361 (1983).\\
$[13]$ N.M. Shaw et al., Phys . Rev. Lett. 63 1342 (1989).\\
$[14]$ H.-J. Behrend et al., Z. Phys . C 41 7 (1988).\\
$[15]$ C. Wendt et al., Phys . Rev. Lett. 58 1810 (1987).\\
$[16]$ O. Adriani et al., Phys . Lett. B 295 371 (1992).\\
$[17]$ M.Z. Akrawy et al., Phys . Lett B 247 448(1990).\\
$[18]$ M. Gronau, C.N. Leung, J.L. Rosner., Phys . Rev. D 29 2539 (1984)\\
$[19]$ Review of Particle Properties, Phys . Rev. D (1992).\\
$[20]$ M.E.Peskin, T. Takeuchi, Phys . Rev. D 46 381 (1992).\\
$[21]$ M. Gronau, C.M. Leung, J.L. Rosner, Phys . Rev. D 29 2539 (1984).\\
$[22]$ T. Kinoshita Phys . Rev. Lett. 2 477 (1959).\\
$[23]$ D.A. Bryman et al., Phys . Rev. Lett. 50 7 (1983).\\
$[24]$ P. Langaker, Lectures given at (TASI '92):Black Holes and
Strings to Particles, Boulder, CO, Bulletin Bd.:
hep-ph@xxx.lanl.gov-9303304 (1993).\\
$[25]$ T. Cheng \& L. Li, Gauge Theory Of Elementary Particle Physics,
Clarendon Press. Oxford (199 ).\\
$[26]$ O. Shanker, Phys . Rev. D, 20 1608 (1979).
$[27]$ A. Blondel et al., Proceedings of the ECFA Workshop on LEP 200
Vol 1 p120 Cern 87 - 08 (1987).\\
\newpage

\section*{Figures}
\begin{description}
\item[Figures {\bf 1}, {\bf 2} and {\bf 3}] show bounds placed on the heavy
neutrino masses by considering all
the constraints in chapters three and four.
The plots are of the neutrino masses $M_{D_i}$ against the ratio
$1/\Lambda_{D_i}=M_i/M_{D_i}$. Any region to the left of a solid line is
forbidden by experiment. The dashed line is the line along which the
masses $M_{D_i}$ are equal to the electron, muon and tau masses.
\newline\\
{\bf Fig 1}
Regions excluded(to the left of curves) in the $M_1$, ${1\over\Lambda_1}$ plane
from: (a)ref[5] Massive
neutrinos in pion decays; (b)ref[9] and (c)ref[8] Massive neutrinos
in Kaon decays; (d)ref[10] Massive neutrinos in D meson decays;
(e)ref[21] /section (4.2) Banching ratio
$\Gamma(\pi\rightarrow e\nu)/\Gamma(\pi\rightarrow \mu\nu)$; (f)ref[16]
section (3.4) $Z$ decays.
\newline\\
{\bf Fig 2}
Regions excluded(to the left of curves) in the $M_2$, ${1\over\Lambda_2}$ plane
from: (a)ref[9] Massive neutrinos
in Kaon decays; (b)ref[10] and (c)ref[10] Massive neutrinos in D meson decays;
(d)Section(4.1.1) Changes in $G_F$, (e)ref[16] section (3.4)
$Z$ decays.
\newline\\
{\bf Fig 3}
Regions excluded(to the left of curves) in the $M_3$, ${1\over\Lambda_3}$ plane
from: (a) section(4.1.2) Tau decays; (b)ref[16] section (3.4)
$Z$ decays.

\item[Figures {\bf 4}, {\bf 5},  {\bf 6} and {\bf 7}]
Regions excluded(to the left of curves) in the $M_2$, ${1\over\Lambda_2}$ plane
from the experimental constraints from chapters three four and five.
Each plot has the same ratio between the masses $M_{D_1}$, $M_{D_2}$ and
$M_{D_3}$ induced by
$D$, This is chosen to be
$M_{D_1}:M_{D_2}:M_{D_3}=M_e:M_{\mu}:M_{\tau}$.
The dashed line correponds to the line along which $M_{D_1}=M_e$,
$M_{D_2}=M_{\mu}$ and $M_{D_3}=M_{\tau}$.
The different plots correspond to three different ratios of $M_1:M_2:M_3$
(the final neutrino masses).
\newline\\
{\bf Fig 4}
Case (1): $M_1=M_2=M_3$.\\
Regions excluded(to the left of curves) in the $M_2$, ${1\over\Lambda_2}$ plane
from: (a)All the restrictions in the $M_1$ ${1\over\Lambda_1}$ plane
studied in chapters three and four; (b)All the restrictions in the
$M_2$ ${1\over\Lambda_2}$ plane studied in chapters three and four;
(c)All the restrictions in the
$M_3$ ${1\over\Lambda_3}$ plane studied in chapters three and four;
(d)Bounds from $\mu\rightarrow e\gamma$;
(e)Bounds from $\mu\,Ti\rightarrow e\,Ti$.
\newline\\
{\bf Fig 5}
Case (2): $M_1:M_2=1:15:60$.\\
Regions excluded(to the left of curves) in the $M_2$, ${1\over\Lambda_2}$ plane
from: (a)All the restrictions in the $M_1$ ${1\over\Lambda_1}$ plane
studied in chapters three and four; (b)All the restrictions in the
$M_2$ ${1\over\Lambda_2}$ plane studied in chapters three and four;
(c)All the restrictions in the
$M_3$ ${1\over\Lambda_3}$ plane studied in chapters three and four;
(d)Bounds from $\mu\rightarrow e\gamma$;
(e)Bounds from $\mu\,Ti\rightarrow e\,Ti$.
\newline\\
{\bf Fig 6}
Case (3): $M_1:M_2:M_3=M_e:M_{\mu}:M_{\tau}$.\\
Regions excluded(to the left of curves) in the $M_2$, ${1\over\Lambda_2}$ plane
from: (a)All the restrictions in the $M_1$ ${1\over\Lambda_1}$ plane
studied in chapters three and four; (b)All the restrictions in the
$M_2$ ${1\over\Lambda_2}$ plane studied in chapters three and four;
(c)All the restrictions in the
$M_3$ ${1\over\Lambda_3}$ plane studied in chapters three and four;
(d)Bounds from $\mu\rightarrow e\gamma$;
(e)Bounds from $\mu\,Ti\rightarrow e\,Ti$.
\item[{\bf 7}] Feynman diagrams for the flavour changing
processes of chapter five.
\end{description}

\end{document}